# Glass Transition in a 2D Lattice Model

E. Eisenberg[1] and A. Baram[2]

[1] Department of Physics, Bar-Ilan University, Ramat-Gan 52900, Israel

[2] Soreq NRC, Yanve 81800, Israel.

**Abstract**:

The dynamics of compaction of hard cross-shaped pentamers on the 2D square lattice is investigated. The addition of new particles is controlled by diffusive relaxation. It is shown that the filling process terminates at a glassy phase with a limiting coverage density $\rho_{rcp} = 0.171626(3)$, lower than the density of closest packing $\rho_{cp} = 1/5$, and the long time filling rate vanishes like $(\rho_{rcp} - \rho(t))^2$. For the entire density regime the particles form an amorphous phase, devoid of any crystalline order. Therefore, the model supports a stable random packing state, as opposed to the hard disks system. Our results may be relevant to recent experiments studying the clustering of proteins on bilayer lipid membranes.

**PACS**: 64.70.Pf, 61.20.Lc, 05.70.Ln

Random filling and compaction processes of systems of hard particles have been of interest as models for the evolution and structure of liquids, glasses, colloidal suspensions and monolayers [1-4]. Although these models have been intensively studied for the last three decades [2], the description of their structure and dynamics still remains an interesting open question of condensed matter science.

One of the primary difficulties in studying dense random systems is that of determining whether the system is truly random. Bernal [1] was one of the first to study experimentally such a system, using plasticene spheres which were compressed together. His results suggested the existence of a limiting density, the *random closest packing* density, which is smaller than the closest packing density. Following Bernal, many workers have investigated, both experimentally and numerically, the possibility of random closest packing state in hard disks and hard spheres systems [2-12].

Mechanical compactions [5,6] and computer densification processes [7] of random configurations of hard disks show that, asymptotically, polycrystalline textures are formed, whose limiting density is the density of closest packing of hard disks. These results support the conjecture of lack of stability for random close packing of hard disks. Numerical simulations of non-spherical systems of particles (regular pentagons and regular heptagons) are less conclusive [8]. The simulations suggest that the densification processes terminate at densities lower than the corresponding densities of closest packing, although small ordered zones appear at these densities. In three dimensions, recent computer simulations of hard sphere systems [9] indicate that the amorphous phase is not stable, and it eventually crystallizes for all densities above the melting density. These results are in contrast with earlier simulations of Lubachevsky et al [7] and Speedy [12], that found dynamically stable amorphous configurations of hard spheres. A recent experimental study on a suspension of hard sphere like particles found the glass to remain stable on earth but to crystallize in the absence of gravity [10].

Another branch of models for generating random configurations is the so-called "deposition models", in which particles are deposited sequentially onto a surface, according to some given rule, up to the point where no additional particle can be deposited (the jamming limit) [4,13]. Usually, the rule is given in terms of the exclusion shell of a particle, which determines which sites in the neighborhood of a given particle are blocked for further depositions. Alternatively, one may define the model by studying extended particles, with hard-core interactions. In general, the random sequential absorption (RSA) models produce dense random systems, but do not have the characteristics associated with systems in which the particles are free to move. This observation motivated the study of diffusive relaxation in RSA models.

The formation of monolayers and two dimensional densification processes [4,14] are frequently dominated by the combined effect of irreversible random deposition of extended particles on a surface (RSA), followed by in-plane diffusion of the deposited particles (RSAD). While the RSA process results in a jammed state with a characteristic jamming limit density $\rho_r$, depending on the details of the interparticle repulsion [13], in-plane diffusion may correct non-effective random depositions, thus allowing for the deposition of additional particles and increasing the density. The structure of the final state is not always obvious. The amorphous phase may crystallize to a crystalline state or a polycrystalline state with final density $\rho(\infty)$ that equals the density of closest packing $\rho_{cp}$. Alternatively, it may retain its amorphous nature with a characteristic final density $\rho(\infty) = \rho_{rcp}$ lower than the closest packing density.

The dynamics of 1D RSAD lattice models is well known [15,16]. In particular, the coverage density $\rho(t)$ converges via a $t^{-1/2}$ power law behavior to its closest packing density. The time dependence of the RSAD process on the two-dimensional square lattice has been investigated by computer simulations for two exclusion models: nearest neighbors exclusion ($N_1$), and nearest neighbors and next nearest neighbors exclusion ($N_2$) [17-19]. For both models the initial amorphous state crystallizes to form ordered regions separated by domain walls. The long time filling is performed through a line by line migration of domain walls, resulting in a unification of neighboring regions. As a result the long time dynamics is expected to be a 1D like, and indeed the density converges to the density of closest packing via a $t^{-1/2}$ power law behavior. For both models stable random closest packing configurations don't exist, in accord with the experimental [5-6] and numerical results [7] found for hard disk systems.

In this letter we present a simple 2D lattice model, which generates a dynamically stable random phase at a density much lower than the closest packing density. This model clearly shows the random closest packing effect. The time evolution of a RSAD filling process on the 2D square lattice is investigated for an exclusion model that extends up to the third shell of neighbors ($N_3$) (the model is identical to the model of hard-core cross shaped pentamers on the square lattice). We show that, in contrast to the $N_1$ model and the $N_2$ model, the RSAD process is unable to correct all local non-effective depositions. As a result the $N_3$ model retains its amorphous nature, and the filling process terminates at a stable density of random closest packing $\rho_{rcp} < \rho_{cp}$. This model describes a RSAD process of more complex, anisotropic objects. Recent experiments have studied the deposition and diffusion of two forms of cytochrome on bilayer lipid membranes. One should expect to see such glass transitions for the deposition of non-isotropic proteins.

Our RSAD model is defined as follows. We start with an empty 2D square lattice containing N=$L^2$ sites, with periodic boundary conditions (L divisible by 5). Particles are deposited at random on the lattice according to the RSA and exclusion rules. Each deposited particle fills a lattice site and excludes further deposition on that site and on its twelve nearest neighbor sites. We study the limit in which the deposition is infinitely fast compared to the diffusion, and therefore the jammed state, whose density is $\rho_r = 0.139750(2)$ [20], is formed in zero time (the figure in parentheses indicate the uncertainty of the last digit). This limit is equivalent to a fast cooling of the system. Then in each time step each particle moves, if possible, with probability $\varepsilon/4$ from its site to one of its four neighboring sites. Migration of a particle is possible only if it does not violate the exclusion rules of the model. After every movement of a particle, new particles are deposited if possible. Vacancies available for additional particles may be formed in five sites that are not excluded any more by the diffused particle. A fruitful motion may result in the deposition of one, or two, or three new particles.

The lattice consists of five sub-lattices that correspond to the five lattice sites covered by a pentamer. It is convenient to label the sub-lattices by an integer n defined by n={(x-2y) modulo 5}, where x and y are the coordinates of a site (note that the lattice may be divided into sub-lattices in two different ways). At any stage of the process, the lattice contains ordered regions, in which the particles are densely packed. Neighboring regions are separated by domain walls, which connect sites that are not covered by pentamers. Only particles that are located close to a domain wall may diffuse to a neighboring site. A movement of a particle results in a transfer of the diffused particle from one sub-lattice to another one. In the RSAD process of the $N_1$ model, the newly added particles are in the same sub-lattice as the diffused particle. Therefore, the filling process corrects local deficiencies, resulting in the formation of growing ordered regions. However, for the $N_3$ model, only two sites out of the five optional vacant sites formed by the diffusion are in the same sub-lattice as the diffused particle. Therefore, in the case of a fruitful diffusion, the newly deposited particles may be added in a non-effective way that increases local disorder. Such non-effective depositions may inhibit additional filling. In addition, local defects generated by inefficient depositions are much more stable in the $N_3$ model due to the "friction" generated by the non-smooth cross shape of the pentamers. An example is given in figure 1, which presents a random closest packing state obtained on a 20x20 lattice. The state is completely frozen, i.e., no particle can move. In the upper part of the lattice a big ordered region can be seen, in which all the particles are in the same sub-lattice. However, pentamer A (marked in the figure) is an example of a non-effectively deposited particle. One can easily see that had it been deposited one site to the left, diffusional relaxation would have resulted in an ordering of its neighborhood. However, once it is deposited in its place, the defect generated is dynamically stable.

We have performed numerical MC simulations of the model, studying a periodic square lattice with length L=1000 sites. The results for the time dependence of the density, averaged over 1000 MC realizations, are presented in Figure 2. The standard deviation in the average value of $\rho(t)$ does not exceed $3 \times 10^{-6}$ for all t, $0 \leq \varepsilon t \leq 5000$. Analysis of the long time behavior points out that the limiting density of the RSAD filling process is a stable random closest packing density $\rho_{rcp} = 0.171626(3)$, clearly lower than the density of closest packing $\rho_{cp} = 1/5$, confirming the above arguments. The density converges to its limiting value via a $t^{-\alpha}$ power law, where $0.90 \leq \alpha \leq 1.15$ is consistent with the numerical results. Assuming $\alpha = 1$, one obtains the following long time expression for the density:

$$\rho(t) = 0.171626(3) - \frac{0.574(1)}{\varepsilon t} + O((\varepsilon t)^{-2}) \qquad (1)$$

The $\alpha = 1$ assumption implies that the filling rate, $d\rho(t)/dt$, vanishes asymptotically like $(\rho_{rcp} - \rho(t))^2$. It should be recalled that, for both $N_1$ and $N_2$ models, the density converges to the density of closest packing via a $t^{-1/2}$, 1D like, power law behavior [19], and the filling rate vanishes asymptotically like $(\rho_{cp} - \rho(t))^3$. This difference manifests the difference in the long time dynamics of the models. The filling progress at long times is related to the probability of two vacancies to get in contact. Therefore, the time dependence is the same as the probability of two independent random walkers to get in contact. For the $N_1$ and $N_2$ models the asymptotic structure is that of large regions, and the density is increased by movements of long lines along the domain walls, resulting in a 1D-like diffusive time dependence. In the present case, however, the asymptotic structure is highly disordered, and the filling progress is related to a 2D random walk process in which small vacancies combine to generate the place for an additional particle.

The density gain $\rho(t) - \rho_r$ is given by an expansion in powers of $\varepsilon t$, whose coefficients are spatial correlation functions at the jamming limit [16]. Using the 1000 MC simulations we computed the first two coefficients in the power series

$$\rho(t) - \rho_r = 0.00991(1)\varepsilon t - 0.00471(5)(\varepsilon t)^2 + O((\varepsilon t)^3) \qquad (2)$$

The deviation of the second order approximation for the density, Eq. (2), from the "exact" (MC) result is less than 5% only for the ultra short time limit, $\varepsilon t \leq 0.5$. At the later stages of the process we approximate the density by the following [2/2] constrained Pade approximant

$$\rho(t) - \rho_r = \frac{0.00991}{u + 0.4754 - .00589/(u + .03581)} \qquad (3)$$

where $u = 1/\varepsilon t$. The four coefficients of the approximant are determined by the first two terms in the short time expansion (eq. (2)) and by the requirement that it fits the long time behavior of eq. (1). In figure 2 the constrained Pade approximant is compared to the "exact" MC numerical results. The fit is excellent for the entire time domain. The maximum relative error, at $\varepsilon t \approx 20$, is of order of 1%.

The contrast between the ordered texture of the $N_1$ model and the amorphous texture of the $N_3$ model is reflected in the contrast between their pair correlation functions as well. In the $N_1$ model the pair correlation function $g_{j,k}(t)$ (the probability that at time t both sites (0,0) and (j,k) are occupied) approaches unity (zero) for even (odd) values of j+k, via the characteristic $t^{-1/2}$ power law. Consequently, the linear size l of a typical ordered crystalline region diverges, in the long time limit, as $\sqrt{\varepsilon t}$. However, for the $N_3$ model, the asymptotic values of the pair correlation functions depend on the position (j,k), thus reflecting its amorphous nature. It is convenient to measure the size of ordered regions by the position of the change of sign of the modified pair correlation function $\rho_{i,j}(t)$ defined by:

$$\rho_{i,j}(t) = (g_{i,j}(t) - \rho(t))/(1 - \rho(t)) \qquad (4)$$

where $-1/4 \leq \rho_{i,j}(t) \leq 1$. Vanishing of this function means lack of correlation, while negative values correspond to anti correlation. Figure 3 presents the time evolution of the modified correlation functions: $\rho_{2,1}(t), \rho_{4,2}(t), \rho_{5,0}(t)$ and $\rho_{6,3}(t)$. All four sites are in the same sub-lattice as the reference (0,0) site (see figure 1). At the jamming limit the modified correlation functions of the sites (4,2), (5,0) and (6,3) approximately vanish, reflecting the lack of even short-range order for t=0. At later stages of the process $\rho_{4,2}(t)$ and $\rho_{5,0}(t)$ increase monotonically as a function of time to their limiting values. On the other hand $\rho_{6,3}(t)$ decreases, although very weakly, with time to its limiting value $\rho_{6,3}(\infty) = -0.0093(1)$, indicating that the site (6,3) is out of the region built around the site (0,0).

The average number of particles with smallest possible distance, $\sqrt{5}$ lattice units, from a given particle $n_{\min}(t) = 8g_{2,1}(t)$ increases from $n_{\min}(0) = 1.912(1)$ at the jamming limit to $n_{\min}(\infty) = 2.824(1)$ at the random closest packing. This contact number is substantially lower than the corresponding contact number of the crystal that equals 4. Furthermore, the average number of particles at a distance $\sqrt{8}$ depends only weakly on time along the RSAD process and is about 0.73. This number reflects the probability of a particle to be along a domain wall.

Our model assumes infinitely fast deposition rate, namely, deposition of a new particle whenever it is possible. This corresponds to infinite cooling of the system, by suddenly reducing the activity to zero. A more realistic model should include the effect of finite cooling, by introducing a finite value for the activity. In that case, deposition of a new particle will occur with a finite probability. The effect of this modification is limited. The asymptotic structure is still amorphous, and a random closest packing state is formed. However, the limiting density increases as the deposition rate decreases.

The lattice gas $N_3$ model is known to undergo a first order phase transition from a low density disordered fluid phase to a high density ordered crystalline phase [21]. The equation of state consists of two different branches separated by a density gap, between $\rho_l \approx 0.16$ and $\rho_r \approx 0.19$. The random closest packing density $\rho_{rcp}$ therefore lies inside the density gap, while the jamming limit is located in the range of the fluid phase. Thus, at the onset of diffusion, the system is ergodic, and only later depositions form the glassy phase. One would expect this glassy phase to occur at a density larger then the fluid transition, and indeed $\rho_{rcp} > \rho_l$. Since the formation of small ordered regions is sufficient to trap the system in its glassy phase, the random closest packing occurs right above $\rho_l$, in the co-existence regime, where such small ordered regions are stable.

In conclusion, a simple lattice model in two dimensions is proposed, which exhibits a glass transition as a result of fast cooling. The asymptotic state has a well-defined density, smaller than the closest packing one, which is a manifestation of the so-called random closest packing. The long time filling rate vanishes like $(\rho_{rcp} - \rho(t))^2$. Spatial correlation functions are studied as a function of time, and it is shown that, for the entire density regime, the particles form an amorphous phase devoid of any crystalline order.

**Figure Captions:**

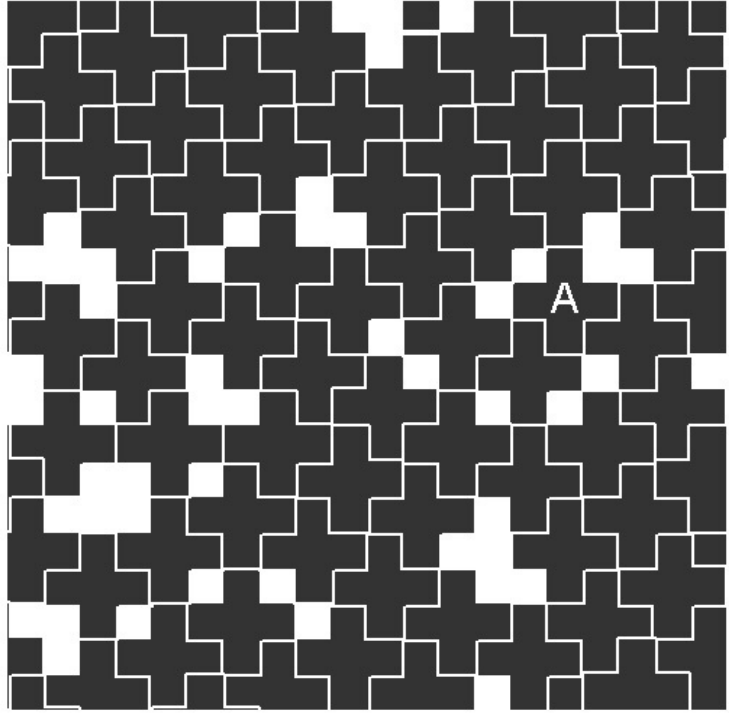

Figure 1

**Figure 1**: An example of a random closest packing state obtained by the RSAD process on a 20x20 lattice with periodic boundary conditions. Note the ordered region in the upper part and the stable defects, which are accompanied by empty sites (white-colored). The density is 70/400=0.175.

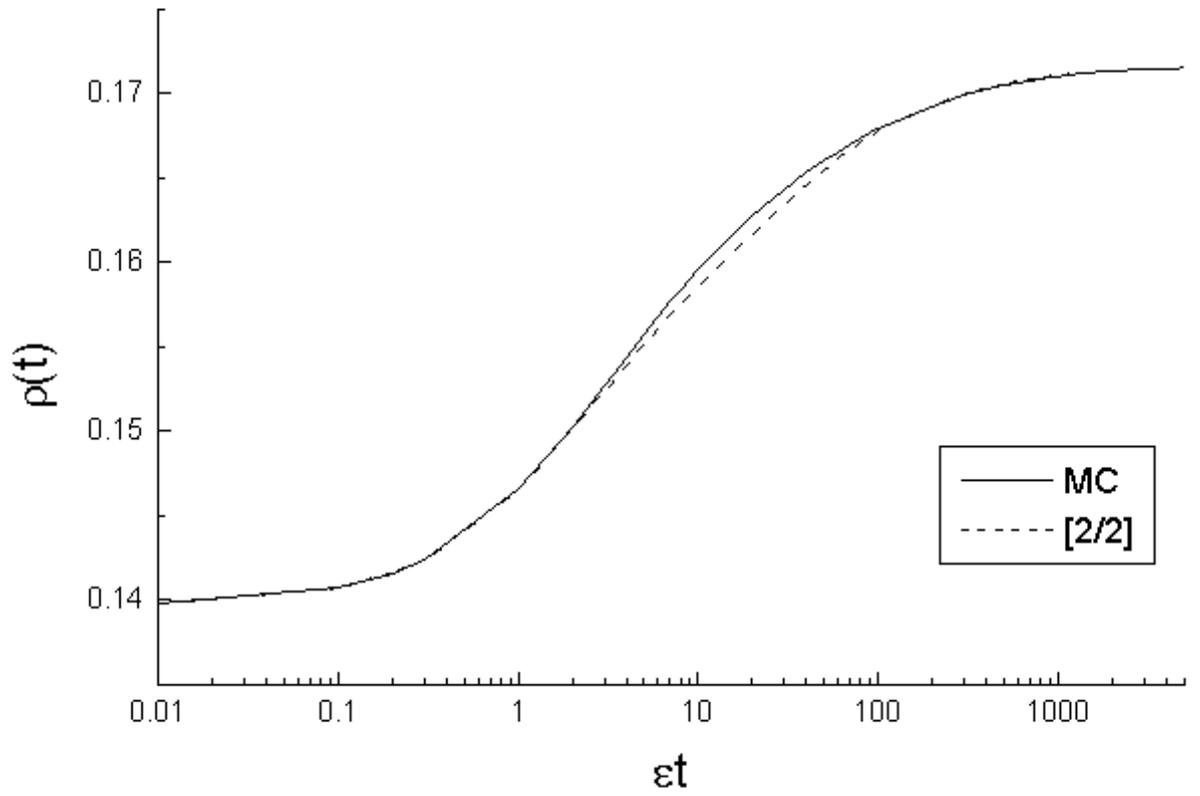

**Figure 2**: Coverage density vs. time, averaged over 1000 MC realizations (solid), compared to the [2/2] Pade approximant (3) (dashed).

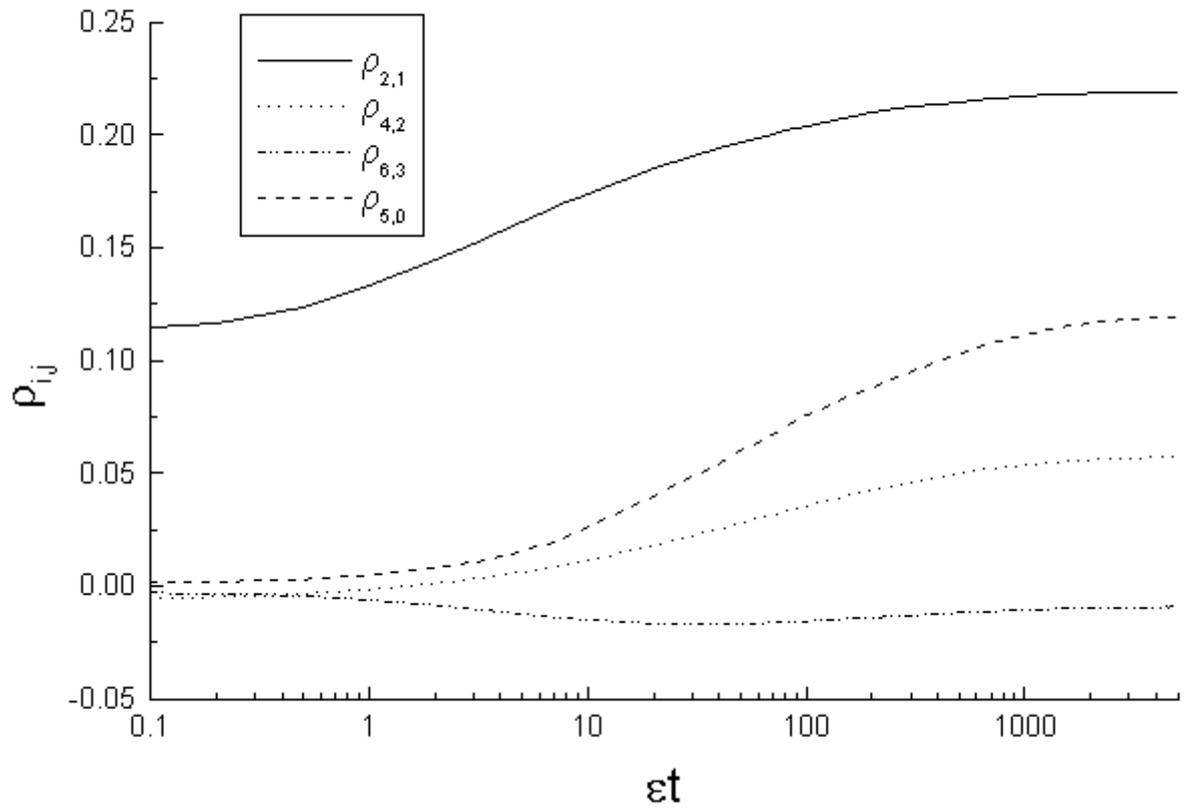

**Figure 3**: Correlation functions as a function of time.